%% file: ratchet_myfile4april.tex
\documentclass[aps,prl,twocolumn]{revtex4-1}
\usepackage{bm}
\usepackage{graphicx}
\usepackage{amsmath}
\usepackage{amssymb}
\usepackage[utf8]{inputenc}
\usepackage[T1]{fontenc}
\usepackage{color}
\usepackage{upgreek}
\usepackage{subfigure}
\usepackage[normalem]{ulem}
\usepackage[unicode=true,colorlinks=true,citecolor=blue]{hyperref}
\usepackage{placeins}
\usepackage{lipsum}
\usepackage{epsfig}
\usepackage[utf8]{inputenc}
\usepackage{wrapfig}
\newcommand{\nix}[1]{}
\definecolor{greenI}{rgb}{0, .4, 0}

\usepackage[version=4]{mhchem}
\usepackage{siunitx}

\newcommand{\MilliMeter}[1]{\SI{#1}{\milli\metre}}
\newcommand{\MilliElectronVolt}[1]{\SI{#1}{\milli\electronvolt}}

\newcommand{\MicroMeter}[1]{\SI{#1}{\micro\meter}}
\newcommand{\NanoMeter}[1]{\SI{#1}{\nano\meter}}

\newcommand{\CentiMeterSquaredPerVoltSeconds}[1]{\SI{#1}{\centi\meter\squared\per{\volt\second}}}
\newcommand{\MilliWatt}[1]{\SI{#1}{\milli\watt}}

\newcommand{\TeraHertz}[1]{\SI{#1}{\tera\hertz}}

\newcommand{\MilliAmperePerWatt}[1]{\SI[per-mode=symbol]{#1}{\milli\ampere\per\watt}}

\newcommand{\Degree}[1]{\SI{#1}{\degree}}

\begin{document}

\title{Giant ratchet magneto-photocurrent in graphene lateral superlattices}

\author{S. Hubmann$^1$, V.~V. Bel'kov$^2$, L.~E. Golub$^2$, V.~Yu. Kachorovskii$^{2,3}$, M. Drienovsky$^1$, J. Eroms$^1$, D. Weiss$^1$,  and S.~D. Ganichev$^1$}

\affiliation{$^1$Terahertz Center, University of Regensburg, 93040 Regensburg, Germany}

\affiliation{$^2$Ioffe Institute, 194021 St. Petersburg, Russia}

\affiliation{$^3$CENTERA Laboratories, Institute of High Pressure Physics, Polish Academy of Sciences PL-01-142 Warsaw, Poland}

\begin{abstract}
We report on the observation of the magnetic quantum ratchet effect in graphene with a lateral dual-grating top gate (DGG) superlattice. We show that the THz ratchet current exhibits sign-alternating magneto-oscillations due to the Shubnikov-de Haas effect. The amplitude of these oscillations is greatly enhanced as compared to the ratchet effect at zero magnetic field. The direction of the current is determined by the lateral asymmetry which can be controlled by variation of gate potentials in DGG. We also study the dependence of the ratchet current on the orientation of the terahertz electric field (for linear polarization) and on the radiation helicity (for circular polarization). Notably, in the latter case, switching from right- to left-circularly polarized radiation results in an inversion of the photocurrent direction. We demonstrate that most of our observations can be well fitted by the drift-diffusion approximation based on the Boltzmann kinetic equation with the Landau quantization fully encoded in the oscillations of the density of states.	
\end{abstract}

%\pacs{
%}
\maketitle

\section{Introduction}
\label{introduction}

The discovery of graphene opened a new research direction in condensed matter physics. The unique optical properties of this material prompted a rapid development of photonics and optoelectronics, see, e.g., Refs.~\cite{CastroNeto2009,Bonaccorso2010,Mueller2010,Echtermeyer2011,Novoselov2012,Grigorenko2012}. These are especially important for applications in the terahertz (THz) range of frequencies, see THz roadmap~\cite{Dhillon2017}  and~\cite{Graham2012,Mittendorff2013,Freitag2013,Ryzhii2014,Cai2014,Mittendorff2015,Otsuji2012,Koseki2016,Bandurin2018,Vicarelli2012,Hartmann2014,Tredicucci2014}. THz-radiation-induced non-linear optical effects, including rectification of  THz/infrared electromagnetic waves, offer a new playground for many intriguing	phenomena in graphene, see, e.g., reviews~\cite{Otsuji2012,Glazov2014,Koppens2014,Low2014,Hasan2016,Ganichev2018,You2018,Rogalski2019,Wang2019}. These phenomena deliver graphene-specific mechanisms of photocurrent generation and provide a basis for the development of novel graphene radiation plasmonic detectors. Such detectors are  compact, tunable by gate voltage and  have already shown fast and sensitive operation in a broad frequency band from sub-THz to infrared, and from ambient- to cryogenic temperatures~\cite{Mittendorff2013,Freitag2013,Ryzhii2014,Cai2014,Mittendorff2015,Otsuji2012,Koseki2016,Bandurin2018,Vicarelli2012,Hartmann2014,Tredicucci2014}.

Among the highly promising radiation detecting mechanisms is the ratchet effect, i.e., the generation of a $dc$ electric current responding to an $ac$ electric field in systems with broken $P$-symmetry. This is one of the most general and fundamental nonlinear phenomena in optoelectronics, for reviews see, e.g.,~\cite{Haenggi2009,Linke2002,Reimann2002,Ivchenko2011,Denisov2014,Bercioux2015,Reichhardt2017,Ganichev2018}. In graphene, the ratchet effect can be obtained in monolayers with asymmetric micro-patterns~\cite{Kiselev2011,Ermann2011,Koniakhin2014}, layers with built-in structure inversion asymmetry (SIA)~\cite{Glazov2014,Jiang2011}  (in this case it is typically called photogalvanic effect~\cite{Ivchenko2011,Weber2008}), short-channel devices, like field effect transistors (FETs) with asymmetric boundary conditions~\cite{Vicarelli2012,Tomadin2013a,Muraviev2013,Spirito2014,Cai2015,Wang2015,Auton2017,Bandurin2018a,Bandurin2018},  as well as in structures with asymmetric grating type of electrodes~\cite{Ganichev2018,Nalitov2012,Otsuji2013,Rozhansky2015,Olbrich2016,Popov2016,Fateev2017,Fateev2019}. Besides their fundamental significance, the two latter types of ratchets are extremely important for applications, since they provide a very promising route towards fast, sensitive, and gate-tunable detection of THz radiation at room temperature. The study of ratchet effects in graphene under different transport regimes such as the drift-diffusion~\cite{Nalitov2012,Olbrich2016} or the hydrodynamic one~\cite{Rozhansky2015,Popov2016,Fateev2017,Fateev2019}, including plasmonic effects, is a very challenging task that has just begun to be explored.

For ratchets new physics comes into play, when an external magnetic field is applied. In recent work we showed, e.g., that a ratchet effect can be induced by an external magnetic field even in case of homogeneous graphene with structure inversion asymmetry~\cite{Drexler2013}. The effect was called magnetic quantum ratchet effect and belongs to the class of magneto-photogalvanic effects~\cite{Falko1989,Ivchenko1988,Belkov2005,Tarasenko2008,Weber2008a,Tarasenko2011,Zoth2014}. It is sensitive to disorder and tunable by a gate voltage~\cite{Drexler2013}. The observation triggered numerous theoretical proposals aimed to enhance and control magnetic ratchet effects in graphene-based systems. In particular, it was predicted that the magnetic ratchet effect can be enormously increased under cyclotron resonance condition and in periodic grating gate structures. 

\begin{figure}
	\centering
	\includegraphics[width=\linewidth]{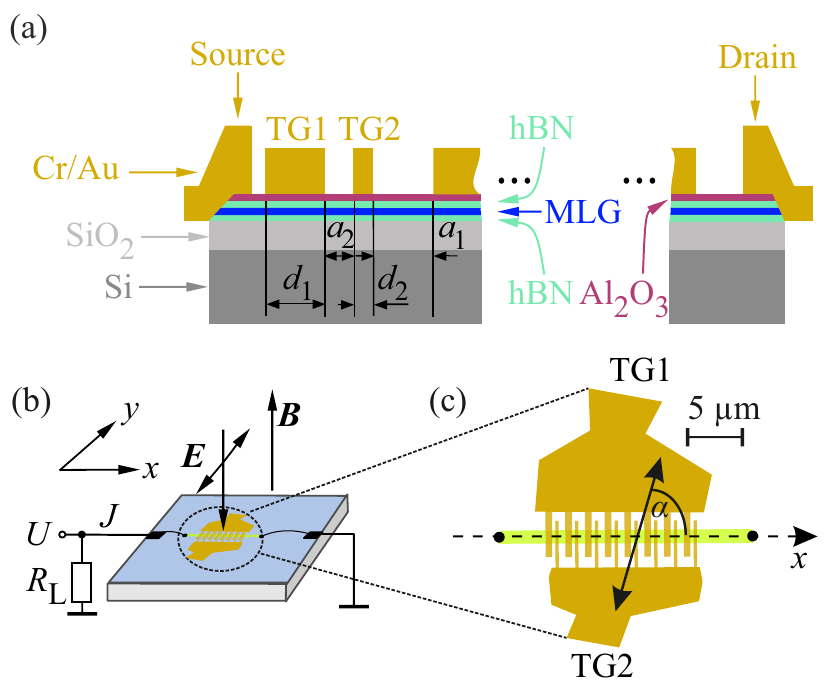}
	\caption{Panel (a): Cross-section sketch of the sample structure and the dual-grating gate superlattice. Panel (b): Sketch of the experimental setup for the ratchet photocurrent measurements. Panel (c): Sketch of the superlattice (top view). 	Black dots show electric contacts to graphene layer.}
	\label{setup}
\end{figure}

Here we report the observation of the giant oscillating magnetic ratchet effect in graphene with superimposed lateral superlattice, consisting of a dual-grating top gate (DGG) structure. We show particularly, that, the THz-radiation-induced ratchet current exhibits sign-alternating magneto-oscillations stemming from Landau quantization and having the same period as the Shubnikov-de Haas (SdH) oscillations. The amplitude of the ratchet current oscillations is greatly enhanced (at least by one order of magnitude) as compared to the ratchet effect at zero magnetic field previously studied in similar structures~\cite{Olbrich2016}. The latter effect was  shown to be caused by the combined action of a spatially periodic in-plane potential and the spatially modulated light due to the near-field effects of the radiation diffraction~\cite{Olbrich2016}. Quantum oscillations appear also as a function of top/back gate voltage in our DGG structures when subjected to a constant magnetic field. Thereby, the direction of the current is controlled by the lateral asymmetry parameter~\cite{Ivchenko2011,Olbrich2016}
	\begin{equation}
	\label{Xi1}
	\Xi = \overline{|\bm E(x)|^2 {dV(x)\over dx}}.
	\end{equation}
Here, the overline stands for the average over the ratchet period, $dV(x)/dx$ is the derivative of the coordinate dependent electrostatic potential $V (x)$, and $\bm E(x)$ the distribution of the radiation electric field being coordinate dependent due to the near-field of diffraction. We show that by changing the individual gate voltages of the dual-grating gate structure we can controllably change the sign of $\Xi$ and, thus, the direction of the ratchet current. Furthermore, we study the response to both linear and circularly-polarized radiation and demonstrate that the magnetic ratchet current is sensitive to the orientation of the linear polarization with respect to the fingers of the DGG structure as well as to the radiation helicity in the case of circular polarization. In the latter case switching from right- to left-circular polarization results in a phase shift of the oscillations by $\pi$, i.e., at constant magnetic field, the current direction reverses. The theoretical modeling of our results is based on the experimental values of all involved parameters and the assumption that the frequency of the incoming radiation is much higher than the plasmonic frequency, so that plasmonic effects do not contribute substantially. We demonstrate that in strong quantizing magnetic fields all observations can be well fitted by the drift-diffusion approximation of the kinetic Boltzmann equation. Within this approach we find that the photocurrent is proportional to second derivative of the longitudinal resistance and, therefore, almost follows the Shubnikov-de Haas  resistance oscillations with a large enhancement factor arising due to differentiating of rapidly oscillating  function. 

The paper is organized as follows. In Sec.~\ref{samples_methods} we describe the investigated samples and experimental technique. In Sec.~\ref{results} we discuss the observed magnetic ratchet effects generated by linearly and circularly polarized THz radiation. In the following Secs.~\ref{theory} and~\ref{discussion} we present the theory 	and compare the corresponding results with the experimental data. Finally, in Sec.~\ref{summary} we summarize the results.

\begin{figure}
	\centering
	\includegraphics[width=\linewidth]{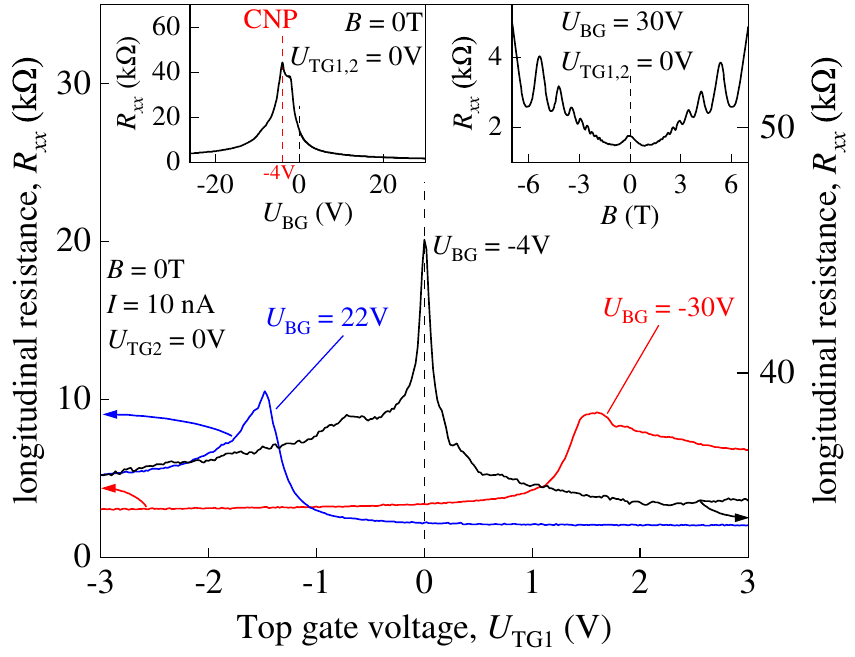}
	\caption{Dependences of the longitudinal resistance $R_{xx}$ on the top gate voltage $U_{\text{TG1}}$ for different back gate voltages. Left and right insets show dependences of $R_{xx}$ on back gate voltage and magnetic field, respectively.
	  }
	\label{transport}
\end{figure}

\section{Samples and methods}
\label{samples_methods}

Single-layer graphene samples encapsulated in hexagonal boron nitride (\ce{hBN}) were prepared using the van-der-Waals stacking technique with Cr/Au edge contacts established by Wang {\em et al.} \cite{Wang2013}. After exfoliation and stacking on top of a silicon wafer with \NanoMeter{285} thermal oxide, a Hall bar mesa was etched using \ce{CHF_3} based reactive ion etching, and edge contacts were deposited by thermal evaporation. To avoid gate leakage at the mesa sidewalls, the samples were covered with \NanoMeter{5} \ce{Al_2O_3} using atomic layer deposition. The highly doped Si wafer serves as a uniform back gate. 

\begin{figure}
	\centering
	\includegraphics[width=\linewidth]{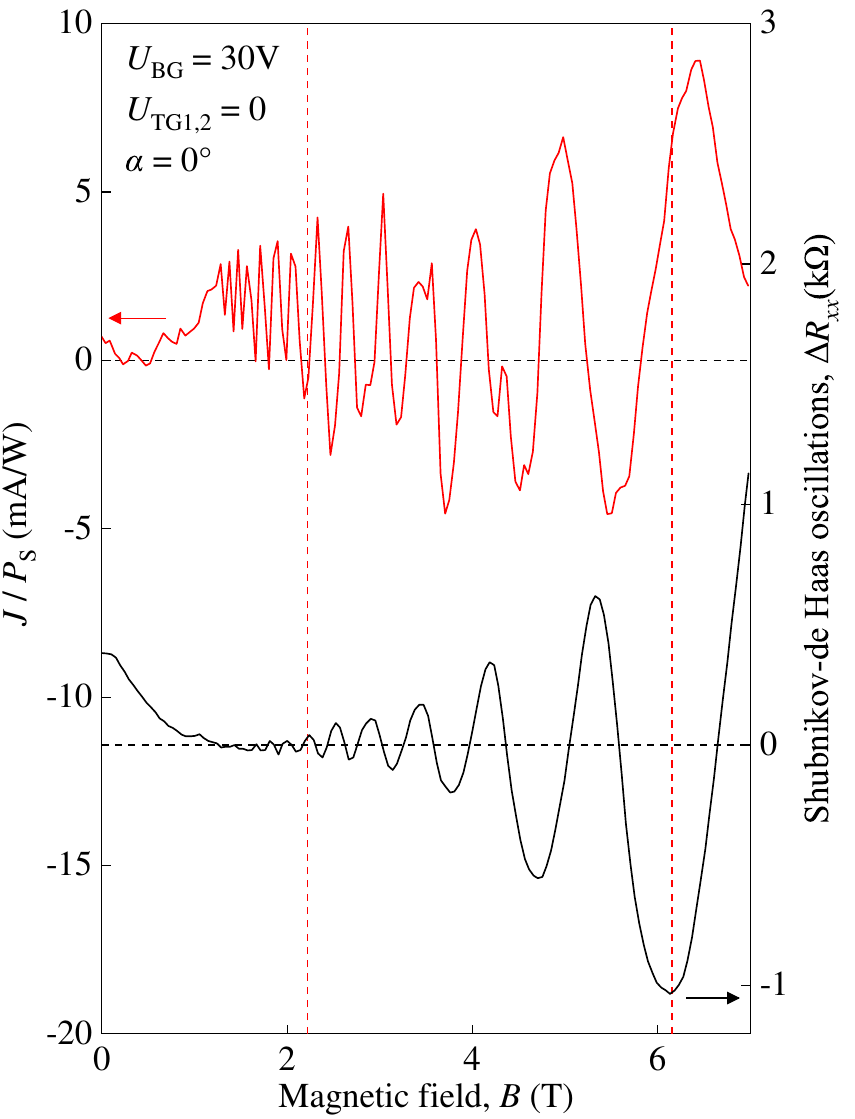}
	\caption{ The photocurrent normalized to the radiation power coming onto the sample, $P_{\rm S}$, and the SdH longitudinal resistance oscillations $\Delta R_{xx}$ as  functions of magnetic field $B$. The oscillatory part $\Delta R_{xx}(B)$ was obtained by subtracting a polynomial background of the form $A_0 + C\,B^2 +D\,B^4$ from the longitudinal resistance $R_{xx}(B)$. The coefficients $A_0$, $C$ and $D$ were obtained by fitting to the high field data.
		}
	\label{magnetic}
\end{figure}

Afterwards, following the recipe of Ref.~\cite{Olbrich2016}, a dual-grating top gate (DGG) superlattice was fabricated on top of the hBN/graphene/hBN flakes for the measurement of the ratchet photocurrent. The micropatterned periodic DGG fingers were made by electron beam lithography and subsequent deposition of metal (\NanoMeter{5} Cr and \NanoMeter{20} Au) on graphene covered by hBN and a \NanoMeter{5} \ce{Al_2O_3} layer. A sketch of this superstructure is shown in Figs.~\ref{setup}(a) and 1(c). Two gate stripes with different widths $d_1=\NanoMeter{600}$ and $d_2=\NanoMeter{300}$ and spacings $a_1=\NanoMeter{600}$ and $a_2=\NanoMeter{300}$ in between form the supercell of the lattice. The cell is repeated eight times resulting in a superlattice with a total length of $\MicroMeter{14.4}$. All wide stripes were connected forming multifinger top gate electrode TG1. Similarly connected narrow stripes formed gate electrode TG2, see yellow areas in Fig.~\ref{setup}(c). Independent bias voltages ($U_{\rm TG1}$, $U_{\rm TG2}$) could be applied to wide and narrow gate stripes making the electrostatic potential asymmetry in the graphene tunable. The width of the whole structure is \SI{1.4}{\micro\meter} yielding the total  area $A=14.4 \times \SI{1.4}{\micro\meter\squared}=\SI{20.2}{\micro\meter\squared}$.

To measure the longitudinal resistance and the photocurrent normal to the DGG stripes Ohmic contacts were fabricated, see Fig.~\ref{setup}(c). Low temperature transport measurements, which are possible in our structure  in  two-point configuration only, showed well resolved SdH oscillations, see right inset in Fig.~\ref{transport}. The oscillatory part $\Delta R_{xx} (B)$ was obtained by subtracting a polynomial background of the form $A_0 + C\,B^2 +D\,B^4$ from the longitudinal resistance $R (B)$. The coefficients $A_0$, $C$ and $D$ were obtained by fitting to the high field data. Also a clear charge neutrality point was observed at $U_{\text{BG}}=\SI{-4}{\volt}$ while tuning the Fermi energy by sweeping the back gate, see left inset of Fig.~\ref{transport}. We obtained an electron mobility $\mu=$ \CentiMeterSquaredPerVoltSeconds{29000} and a hole mobility of $\mu=\CentiMeterSquaredPerVoltSeconds{15000}$ at $T = \SI{4.2}{\kelvin}$. Application of  a positive top gate potential shifts the peak of the longitudinal resistance to higher negative $U_{\text{BG}}$ while for high negative top gate voltages it moves to positive $U_{\text{BG}}$, see Fig.~\ref{transport}.

%Hallo Philipp,
%
%Ich habe die Probe heute noch gebondet. War gar nicht so leicht auf diese Stufe hochzubonden und ich hoffe, dass etwaige Berührstellen der Drähte mit dem Substrat keine Kurzschlüsse verursachen (ansonsten abreiße und neu bonden)
%Ich bin ab Montag zwei Wochen weg, aber du findest die fertige Probe im großen Exikator im Bonder-labor - Petrischale mit Aufschrift Martin D. - A88 - 19.08.16
%Hier die Eckdaten:
%A88
%oberes Al2O3 = 5nm
%oberes hBN = 14 nm
%Graphen: Monolayer
%unteres hBN = 48 nm
%unteres SiO2 = 285 nm
%
%Kontaktmaterial 5 nm Cr, 80 nm Au
%DGG top gate Material: 5 nm Cr, 20 nm Au
%DGG parameter:
%d1 = 600 nm
%d2 = 300 nm
%a_supercell = 1800 nm
%N_perioden = 8
%
%Breite Struktur (erfragt) = 1400 nm
%
%Viel Glück beim messen.
%
%Martin 

As a radiation source for our experiments a continuous wave methanol terahertz laser with a radiation frequency of $f=\TeraHertz{2.54}$ ($\hbar\omega=\MilliElectronVolt{10.5}$) and a radiation power $P$ of the order of \MilliWatt{50} was used \cite{Kvon2008,Ganichev2009,Olbrich2013}. The radiation was focused onto the sample using an off-axis parabolic mirror resulting in a spot size of $\approx$ \MilliMeter{1.3}, which yields an intensity $I\approx\SI{3.8}{\watt\per\centi\meter\squared}$. The radiation power coming onto the sample is calculated after $P_{\rm S}=I \cdot A$. The laser beam had an gaussian shape as checked by a pyroelectric camera \cite{Ganichev1999,Ziemann2000}. The radiation was modulated at about \SI{75}{\hertz} by an optical chopper in order to use standard lock-in technique. 

The optically pumped molecular laser used here emits linearly polarized radiation. In our setup its polarization plane is oriented along the $x$-axis being normal to the dual-grating gate stripes.	In experiments with linearly polarized radiation, the orientation of the radiation electric field vector $\bm E$ was varied by rotation of a mesh grid polarizer mounted behind the quarter-wave plate providing circularly polarized radiation. The azimuth angle $\alpha$ is the angle between the radiation electric field vector and the $x$-direction, Fig. 1(c). The Stokes parameters describe the degree of the linear polarization in the basis ($x$, $y$) and the basis ($\tilde x$, $\tilde y$) rotated by \Degree{45} $P_{\rm L}$ and $\tilde P_{\rm L}$, respectively. In this setup, they are given by
\begin{equation}\label{lin}	
	P_{\rm L}(\alpha) = \cos 2\alpha,\quad   \tilde P_{\rm L}(\alpha) = \sin 2\alpha\:.	
\end{equation}	
In experiments with elliptically polarized radiation, the radiation helicity was varied by rotating the quarter-wave plate by the angle $\varphi$. By that, at $\varphi$ = 0 radiation is linearly polarized and $\bm E$ is parallel to $x$, whereas at $\varphi = \Degree{45}$ and $\varphi= \Degree{135}$ the radiation is circularly polarized  with opposite helicities. In this setup the Stokes parameters are given by~\cite{Belkov2005,Weber2008}
	\begin{equation}\label{circ}
	\begin{aligned}
	P_{\rm L}(\varphi) &= (\cos 4\varphi + 1)/2 ,&\quad \tilde P_{\rm L}(\varphi) = \sin 4\varphi/2  ,\\  P_{\rm circ} &= \sin 2\varphi\:, 
	\end{aligned} 
	\end{equation}
where $P_{\rm circ}$ defines the degree of circular polarization. The ratchet photocurrents were measured in a magneto-optical cryostat at a temperature of \SI{4.2}{\kelvin} as a voltage drop along a load resistor of $R_{\text L}=\SI{100}{\ohm}$ using standard lock-in technique and then calculated using $J=U/R_{\text L}$. In all graphs, the photocurrent is normalized to the radiation power coming onto the sample, $P_{\rm S}$. An external magnetic field with $B$ up to \SI7{\tesla}  is applied normal to the graphene plane, as sketched in Fig.~\ref{setup}(b).

\begin{figure}
	\centering
	\includegraphics[width=\linewidth]{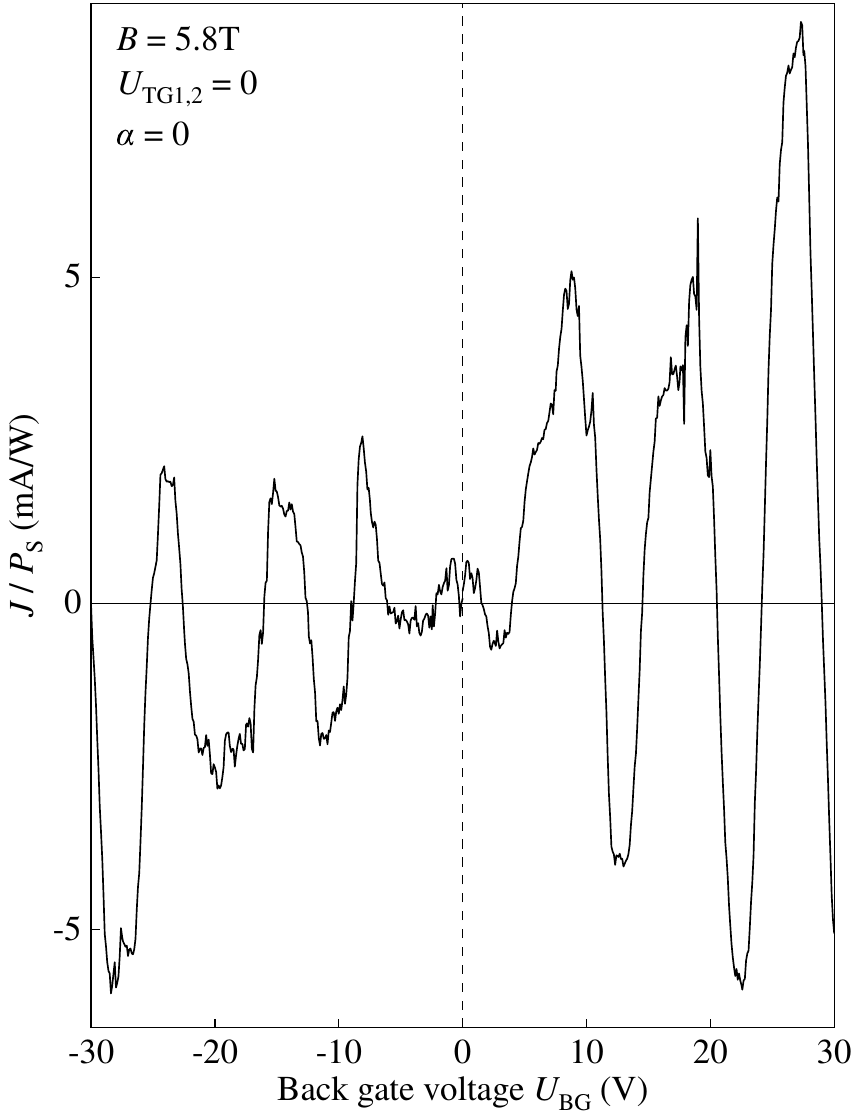}
	\caption{Dependence of the ratchet photocurrent  normalized to the radiation power coming onto the sample, $P_{\rm S}$, on the voltage applied to the back gate. The data are obtained for linear polarization with $\alpha=0$.	}
	\label{backgate}
\end{figure}

\section{Results}
\label{results}

Before discussing our  results on magnetic current we briefly address the photocurrents detected at zero magnetic field. In our previous work~\cite{Olbrich2016}, in which we studied similar structures and applying terahertz radiation with the same parameters, we demonstrated that illumination of the DGG superlattice on graphene results in a photocurrent  exhibiting characteristic behavior of the ratchet effect. In particular, photocurrent direction and magnitude:  (i) are sensitive to the orientation of the radiation electric field vector $\bm E$ and/or the radiation helicity; (ii) depend on the carrier charge sign (electrons/holes);  (iii) are controlled by the lateral asymmetry parameter $\Xi$, which can be varied by applying voltages $U_{\rm TG1}$ and $U_{\rm TG2}$ to the individual subgates. Note that, switching of the gate voltage from $U_{\rm TG1} > 0$, $U_{\rm TG2} = 0$ to $U_{\rm TG1} = 0$, $U_{\rm TG2} > 0$ leads to a change in the sign of $\Xi$  and, as a consequence, to a reversal of the photocurrent direction. Theoretical analysis carried out in Ref.~\cite{Olbrich2016} reveals that the photocurrent is caused by a combined action of a spatially periodic in-plane electrostatic potential and the radiation spatially modulated due to the near-field effects of the diffraction on the DGG stripes. Experiments and theory of this effect present a self-consisted detailed picture of the ratchet current formation, therefore, in the present paper we focused on the magnetic ratchet effect in graphene.

\begin{figure}
	\centering
	\includegraphics[width=\linewidth]{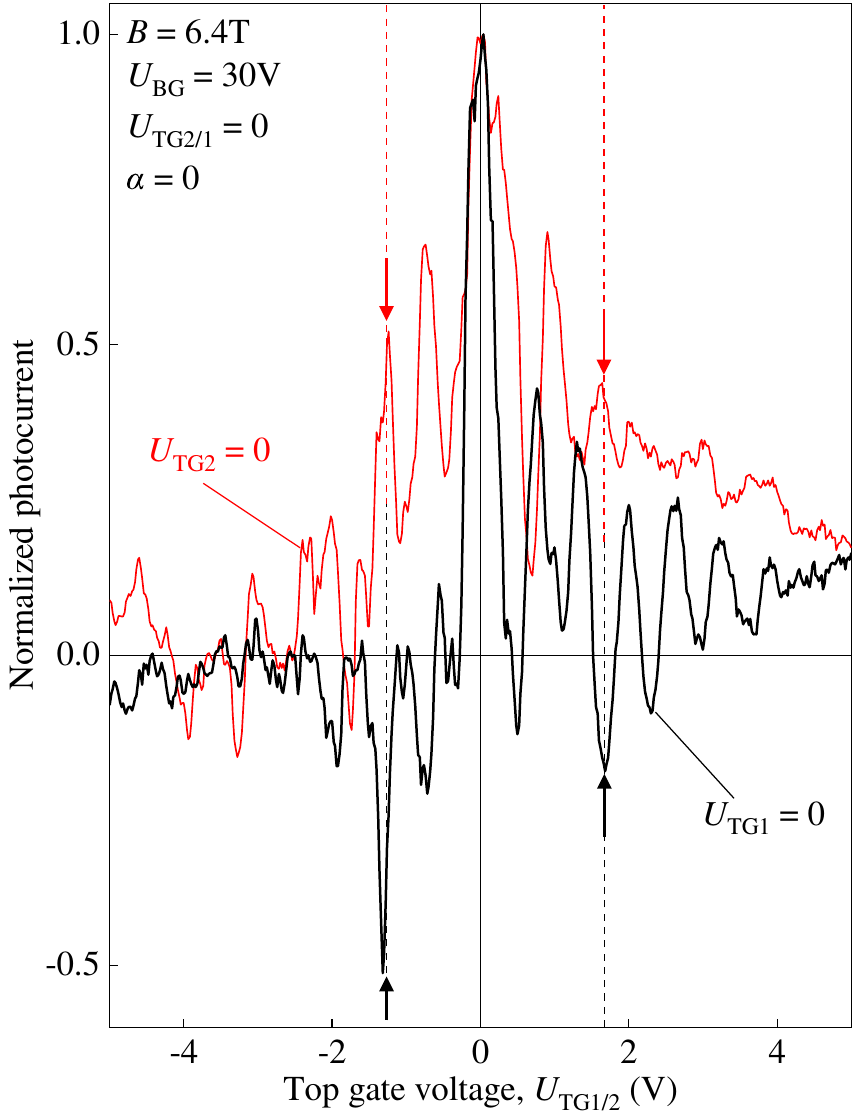}
	\caption{ Top gate voltage $U_{\rm TG1}(U_{\rm TG2})$ dependencies of the photocurrent normalized to the photocurrent maximum. The curves were obtained varying $U_{\rm TG1}$($U_{\rm TG2}$) holding zero bias at the other subgate TG2 (TG1) and for linear polarization with $\alpha=0$. Since at high gate voltages maxima of one dependence correspond to minima at the other this provides an evidence for the ratchet effect being proportional to $\Xi$. This is illustrated by the vertical dashed lines at high positive (negative) gate voltages at which the lateral asymmetry introduced by the applied voltages is stronger than the built-in one caused by the metal stripes deposited on top of graphene.}  
	\label{topgate}
\end{figure}

Applying an external magnetic field we observed that the ratchet photocurrent drastically changes: The photocurrent exhibits sign-alternating SdH-like magneto-oscillations with an amplitude by more than an order of magnitude larger than the photocurrent at zero magnetic field. A characteristic magnetic field dependence is shown in Fig.~\ref{magnetic} for $U_{\text{TG1,2}}=0$, radiation electric field oriented perpendicular to the gate stripes and back gate voltage $U_{\text{BG}}$ = \SI{30}{\volt}. Note that at zero top gate voltages, the lateral asymmetry is created by the built-in potential caused by the metal stripes deposited on top of graphene. 

Comparing the magnetic-field dependencies of the ratchet photocurrent and the longitudinal resistance $R_{xx}(B)$ we found out that  extrema positions of the photocurrent and the SdH oscillations coincide in weak fields. This is  seen clearly in Fig.~\ref{magnetic}, where the left vertical dashed line exemplary indicates extrema positions of the photocurrent and the SdH oscillations. It should be noted that, whereas at low magnetic fields the photocurrent oscillations follow $R_{xx}(B)$, at high magnetic fields a magnetic field-dependent phase shift is present, see right vertical dashed line in  Fig.~\ref{magnetic}. These fields are slightly higher than the magnetic field of the cyclotron resonance ${B_{CR}=\omega \varepsilon_\text{F}/(|e|v_0^2)}$, where $\omega$, $\varepsilon_\text{F}$, and $v_0$ are the radiation angular frequency $\omega=2\pi f$, Fermi energy, and the Dirac velocity in graphene, respectively. Note that  for the carrier density $n_s \approx \SI{E16}{\per\meter\squared}$,  and radiation frequency $f=\TeraHertz{2.54}$, relevant to experiment, we estimated $B_{CR}\approx \SI{1.8}{\tesla}$.

\begin{figure}
	\centering
	\includegraphics[width=\linewidth]{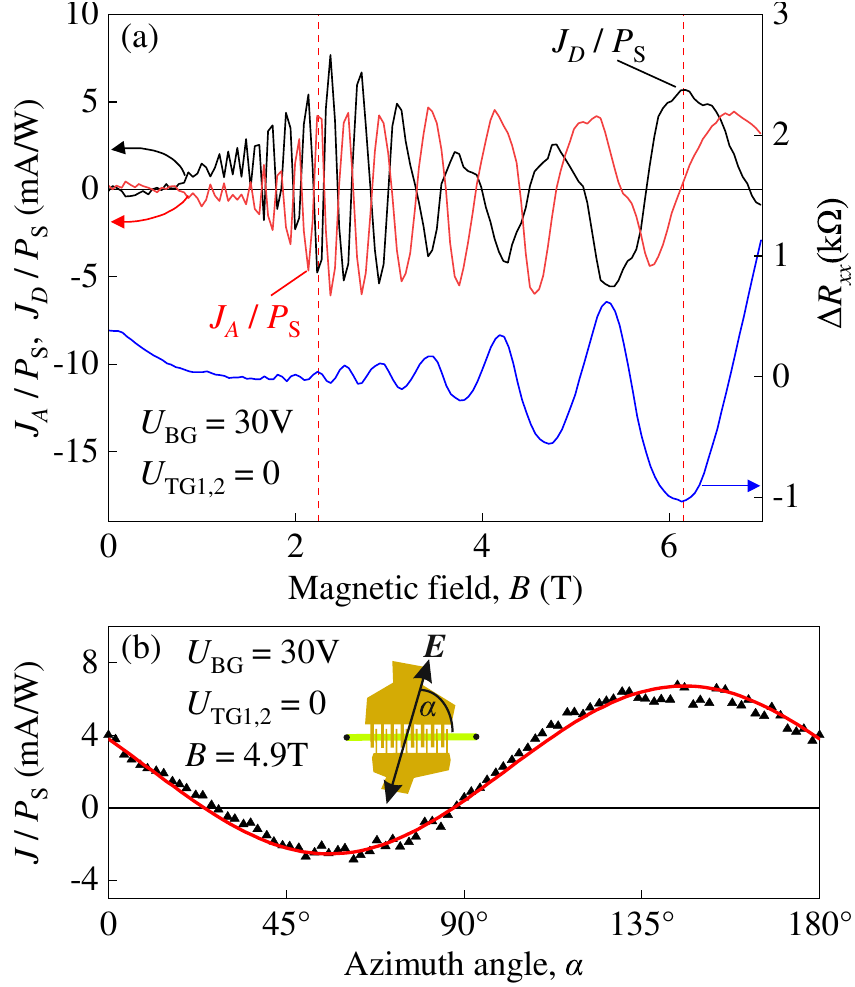}
	\caption{ Panel (a): Magnetic field dependences of the ratchet photocurrent amplitudes $J_A$ and $J_D$ normalized by the radiation power coming onto the sample, $P_{\rm S}$. To extract $J_A$ and $J_D$ we measured magnetic field dependencies for two angles $\alpha =0$ and $\Degree{90}$, at which $P_{\rm L} =\pm 1$ and $\tilde{P}_{\rm L} = 0$. Then, the curves were calculated after $J_A = [J(\alpha=0) - J(\alpha=\Degree{90})]/2$ and  $J_D = [J(\alpha=0) + J(\alpha= \Degree{90})]/2$, where $J(\alpha=0)$ and $J(\alpha=\Degree{90})$ are photocurrent measured for radiation electric field vector oriented $\bm E$   perpendicular and parallel to the DGG stripes, respectively. 
		%Note that while the data, as that presented in the next panel, are obtained for negative magnetic fields  and for visibility they were plotted against  absolute value of $B$. 
		Panel (b):  Dependence of the ratchet photocurrent normalized by the radiation coming onto the sample, $P_{\rm S}$, on the azimuth angle $\alpha$ obtained for a magnetic field of $B=\SI{4.9}{\tesla}$. Red line show fit according to Eq.~\eqref{j_alpha} with fitting parameters $J_A /P_{\rm S}=\MilliAmperePerWatt{1.7}$, $J_{B}/P_{\rm S}=\MilliAmperePerWatt{-4.3}$, and $J_D/P_{\rm S}=\MilliAmperePerWatt{2.1}$.   Inset shows experimental setup and defines angle $\alpha$ describing relative orientation of the radiation electric field vector $\bm E$ and DGG structure.
	}
	\label{alpha}
\end{figure}

The resistance oscillations can also be obtained at fixed magnetic field by the variation of the carrier density $n_s$, e.g. changing the back gate voltage ($n_s \propto U_{BG}$). This kind of oscillations we also observed in the ratchet photocurrent. Figure~\ref{backgate} shows an example of such oscillations obtained for $B =\SI{5.8}{\tesla}$,   $U_{\text{TG1,2}}=0$, and for linear polarization with $\alpha=0$. 

Now we turn to the results obtained by variation of the lateral asymmetry applying different voltages to the top subgates TG1 and TG2.  Figure~\ref{topgate} shows  magnetic ratchet photocurrent oscillations as a function of top gate voltage $U_{\rm TG1}$ ($U_{\rm TG2}$) obtained for zero biased top gate TG2 (TG1), and for linear polarization with $\alpha=0$.  At zero top gate voltages the photocurrent has the same sign and almost the same amplitude. Sweeping voltage of one of the top gates while holding the other one at zero bias we observed that the photocurrent oscillates in a similar way as it is detected for the variation of back gate voltage. However, the period of oscillations is substantially decreased, which clearly follows from the different separation between top/back gates and graphene. At high gate voltages corresponding to  stronger lateral asymmetry as the built-in one we observed that maxima (minima) of the dependence on $U_{\rm TG1}$ corresponds to minima (maxima) of the dependence on $U_{\rm TG2}$. This is illustrated in Fig.~\ref{topgate} for positive and negative top gate voltages by vertical red/black dashed lines. It reveals that the change of sign  of the lateral asymmetry parameter $\Xi$ results in the change of the oscillations sign, as expected for the ratchet effect. Indeed, e.g., for the top gate voltage combinations marked by the right vertical dashed lines ($U_{\rm TG1} > 0$, $U_{\rm TG2} = 0$, red curve, and $U_{\rm TG1} = 0$, $U_{\rm TG2} > 0$, black curve) the signs of the asymmetry parameters $\Xi$ are opposite.

All results discussed previously were obtained for linearly polarized radiation with the electric field $\bm E$ perpendicular to the DGG stripes. Further experiments demonstrate that magneto-oscillations of the ratchet current are sensitive to the  orientation of the THz electric field vector, see Fig.~\ref{alpha}. Our measurements demonstrate that dependence  of the current on the direction of the linear polarization can be well fitted as follows
\begin{equation}
\label{j_alpha}
J=J_AP_{\rm L}(\alpha)+J_B\tilde P_{\rm L}(\alpha)+J_D,
\end{equation}
where $J_A$, $J_B$, and $J_D$ are magnetic field-dependent fitting parameters. Figure~\ref{alpha}(b) exemplary  shows the polarization dependence of the total photocurrent measured at fixed magnetic field $B = \SI{4.9}{\tesla}$. The magnetic field dependence of the coefficients $J_A$ and $J_D$ yielding dominating contributions at low magnetic field are presented in  Fig.~\ref{alpha}(a).  The curves in Fig.~\ref{alpha}(a) were obtained from the magnetic field dependence of the photocurrent excited by the THz electric field vectors $\bm E$ oriented perpendicular ($\alpha=0$) and parallel ($\alpha=\Degree{90}$) to the stripes. For these angles the photocurrent contribution $J_B\tilde{P}_{\rm L}$ is zero and the total current is given by $J = \pm J_A + J_D$. Consequently, the magnetic field dependencies of $J_A$ and $J_D$ were calculated, respectively,  as a half-difference and half-sum  of the photocurrents  measured for $\alpha=0$ and $\Degree{90}$. Figure~\ref{alpha}(a) reveals that these ratchet current contributions have opposite signs and close magnitudes.

\begin{figure}
	\centering
	\includegraphics[width=\linewidth]{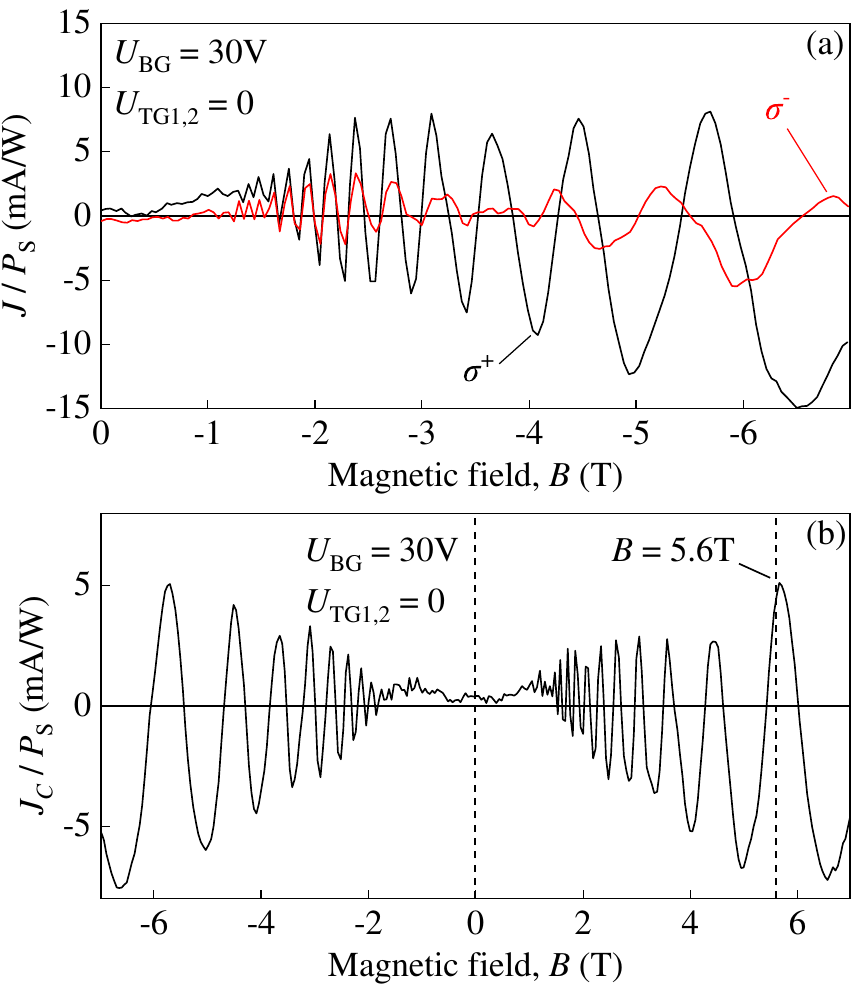}
	\caption{Panel (a): Dependences of the normalized ratchet photocurrent induced by right handed (black curve) and left-handed (red curve) circularly polarized radiation on the magnetic field $B$. Panel (b): Magnetic field dependence of the amplitude of the  helicity-dependent ratchet current $J_{C}$ normalized by the radiation power $P_S$. $J_{C}$ was calculated according to Eq.~\eqref{JC}. }
	\label{circular}
\end{figure}

\begin{figure}
	\centering
	\includegraphics[width=\linewidth]{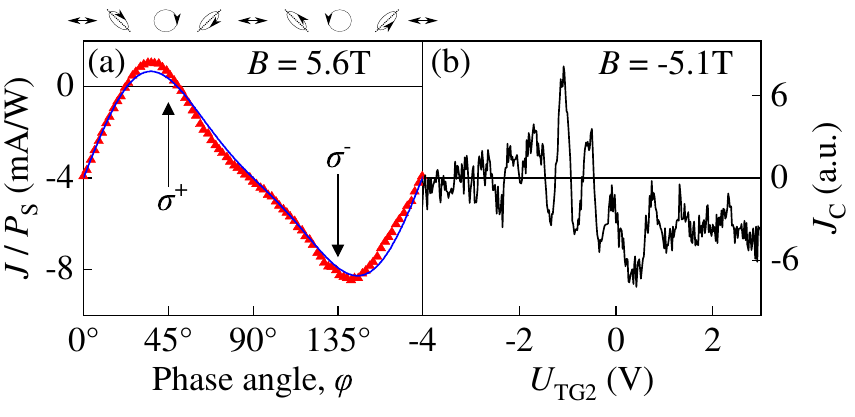}
	\caption{ Panel (a): Dependence of the ratchet photocurrent  on the radiation helicity for a magnetic field of $B~=~\SI{5.6}{\tesla}$. Arrows above the plot illustrate the polarization state. Blue line shows fit according to Eq.~\eqref{J_exp} with fitting parameters $J_{A}/P_{\rm S}=\MilliAmperePerWatt{-0.2}$, $J_B/P_{\rm S}=\MilliAmperePerWatt{1.8}$, $J_C/P_{\rm S}=\MilliAmperePerWatt{4.1}$,   and $J_{D}/P_{\rm S}=\MilliAmperePerWatt{-3.8}$. Panel (b): Dependence of the amplitude of the helicity-dependent ratchet current $J_{C}$ on the second top gate voltage at $B=\SI{-5.1}{\tesla}$.}
	\label{circular2}
\end{figure}

Above we discussed experiments with linear polarization rotated by  $\lambda/2$ plate. Let us now discuss experimental data obtained by using  $\lambda/4$ plate  which  allows us to create circular polarization.  Figure~\ref{circular}(a) shows magneto-oscillations of the photocurrent excited by right- ($\sigma^+$) and left-handed ($\sigma^-$) circularly polarized radiation. Subtracting these two curves we obtain the amplitude of the helicity-sensitive photocurrent $J_{C}$ given by 
\begin{equation}\label{JC}
J_{C}=\frac{J(\sigma^+)-J(\sigma^-)}2.
\end{equation}
This treatment extracts the photocurrent contribution whose direction reverses upon switching the radiation helicity. Note that for circularly polarized radiation the degrees of linear polarization $P_{\rm L}$ and $\tilde P_{\rm L}$ are equal to zero and, consequently, the magnetic ratchet effect caused by the linearly polarized radiation vanishes. This sign-reversion has been observed directly by measuring the dependence of the photocurrent on the phase angle $\varphi$ defining the radiation helicity after Eq.~\eqref{circ}. This is shown in Fig.~\ref{circular2}(a), where the helicity dependence of the ratchet photocurrent is studied at $B=\SI{5.6}{\tesla}$, corresponding to the maximum of $J_{C}$, see Fig.~\ref{circular}(b). The overall polarization dependence of the photocurrent can be well fitted by 
\begin{equation}
\label{J_exp}
J= J_AP_{\rm L}(\varphi)+J_B\tilde{P}_{\rm L}(\varphi)+J_CP_{\rm circ}(\varphi)+J_D,
\end{equation}
where $J_A$, $J_B$, $J_C$, and $J_D$ are magnetic field-dependent fitting parameters. Figure~\ref{circular2}(a) demonstrates that the circular photocurrent yields substantial contribution to the total photocurrent.
Similar to the linear ratchet effect the circular photocurrent  shows clear oscillations upon variation of the gate potential, see Fig.~\ref{circular2}(b) for $J_C(U_{\rm TG2})$.

To summarize the experimental results, we observed that excitation of the DGG superlattice with THz radiation results in the ratchet photocurrent showing magneto-oscillations as well as oscillations upon variation of back/top gate voltages. The oscillations are closely related to the Shubnikov-de Haas effect. Measurements with controllable variation of the top gate voltages and, correspondingly, the lateral asymmetry parameter $\Xi$ clearly demonstrate that the photocurrent is caused by the ratchet effect. The photocurrent is giantly enhanced in the presence of  magnetic field. The experimental results show a substantial contribution of both, linear and circular magnetic ratchet effects exhibiting sign-alternating magneto-oscillations.

\section{Theory}
\label{theory}

In this section we generalize the theory of magnetic ratchets to graphene-based  systems in the  Shubnikov-de-Haas regime. While for ratchets based on  2D systems with a parabolic energy spectrum the theory of magneto-oscillations  was developed in  Refs.~\cite{Budkin2016a,Faltermeier2017,Faltermeier2018} it can not be applied to graphene. Moreover, as it has been  demonstrated in Ref.~\cite{Nalitov2012},  even at zero magnetic field the ratchet currents   are  drastically different in systems with linear and parabolic energy dispersions. 

We use the Boltzmann kinetic equation approach, where the electric current density is given by the following expression 
\begin{equation}
\label{j}
\bm j = e \sum_{\nu, \bm p} \bm v_{\bm p} \bar{f}_{\bm p}.
\end{equation}
Here $\bm v_{\bm p} = v_0 \bm p/p$ is the velocity of carriers having the momentum $\bm p$ with $v_0$ being the Dirac fermion velocity, $\nu$ enumerates spin and valley-degenerate states, and $\bar{f}_{\bm p}$ is the  distribution function $f_{\bm p}(x)$ averaged over the space period of DGG structure. The latter is a solution of the kinetic equation~\cite{Budkin2016a}
\begin{equation}
\label{kin_eq}
\left( {\partial \over \partial t} + v_{\bm p, x}{\partial \over \partial x} + \bm F_{\bm p} \cdot {\partial \over \partial \bm p} \right) f_{\bm p}(x) = \text{St}[f_{\bm p}(x)].
\end{equation}
Here $\text{St}[f]$ is the elastic scattering collision integral, and the space- and time-dependent force is given by 
\begin{equation}
\bm F_{\bm p} = e\bm E(x) \text{e}^{-i\omega t}+e\bm E^*(x) \text{e}^{i\omega t} + {e\over c}\bm v_{\bm p}\times \bm B - {dV\over dx} \hat{\bm x},
\end{equation}
where $\bm E(x)$ is the radiation near-field acting on 2D carriers, $\omega$ is the radiation frequency, and $V(x)$ is the periodic potential of the ratchet.
The distribution function is found by sequential iterations of the kinetic equation in small electric field amplitude and the ratchet potential with the result linear in $dV/dx$ and quadratic in  $|\bm E(x)|$ with the ratchet current proportional to the asymmetry parameter $\Xi$ given by Eq.~\eqref{Xi1}. 

\begin{figure}
	\centering
	\includegraphics[width=\linewidth]{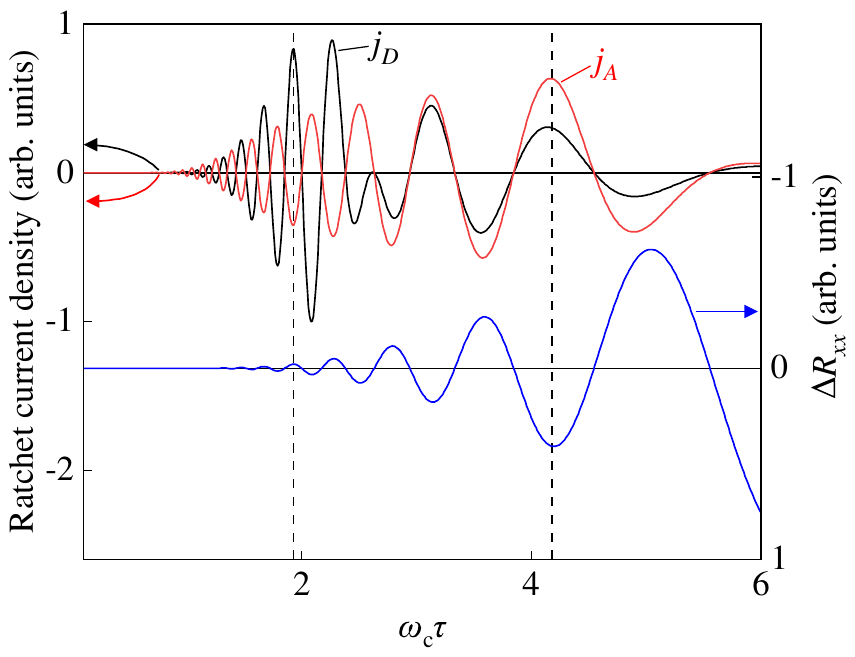}
	\caption{Theoretical dependences of the magnetic ratchet photocurrents on  $\omega_c \tau \propto B$.  Black curve shows the polarization independent contribution $j_D$ ($J_D$ in experiment), red curve is the $P_\text{L}$-linear contribution $j_A$ ($J_A$ in experiment). The photocurrents are calculated by using Eq.~\eqref{j_tot_fin} at $\omega\tau=4.6$, $\tau_q/\tau=0.3$, $\varepsilon_\text{F}\tau/\hbar=25$.
}
	\label{fig:theory}
\end{figure}

For zero  magnetic field the polarization-dependent ratchet current density in graphene is given by~\cite{Nalitov2012,Olbrich2016}

\begin{align}
\label{zerocurrent}
j_x^0=-j_0 {(\omega\tau)^2 \over 1+(\omega\tau)^2} \left[ {2(\omega\tau)^2 \over 1+(\omega\tau)^2}+ P_{\rm L} \right], 
\\ j_y^0=-j_0 {\omega\tau \over 1+(\omega\tau)^2} \left[\omega\tau\tilde{P}_{\rm L}+ {(\omega\tau)^2-1 \over 1+(\omega\tau)^2}P_\text{circ}\right], \nonumber
\end{align}
where $j_0$ is given by: 
\begin{equation}
\label{j0}
	j_0=\Xi {e^3 v_0^2\tau \over 2\pi \hbar^2\varepsilon_\text{F}\omega^2},
\end{equation}
and $\tau$ is the electron elastic scattering time assumed to be independent of the Fermi energy.

Here we calculate the ratchet current in graphene in the Shubnikov-de-Haas regime. The quantization of the energy spectrum in strong magnetic fields is taken into account by the oscillating density of states at the Fermi energy: $g=g_0(1+\delta_c)$, where $g_0=2\varepsilon_\text{F}/(\pi\hbar^2v_0^2)$ is the zero-field density of states with account for spin and valley degeneracies, and the oscillating part is given by~\cite{Zoth2014,Briskot2013}
\begin{equation}\label{delta_c}
\delta_c=2\cos{(\pi\varepsilon_\text{F}/\hbar\omega_c)}\exp{(-\pi/\omega_c\tau_q)}.
\end{equation}
Here the cyclotron frequency in graphene is ${\omega_c=eBv_0^2/\varepsilon_\text{F}}$, and $\tau_q$ is the quantum lifetime. As a result of the density-of-states oscillations, the electron scattering rate also has an oscillating part: $\gamma=\tau^{-1}(1+\delta_c)$.

In Appendix~\ref{App_theory}, we find the distribution function $\bar{f}_{\bm p}(x)$. Note that not only the angular-independent part of the distribution function but also its second angular harmonics contribute to the current Eq.~\eqref{j} in graphene due to non-parabolicity of the Dirac fermion dispersion~\cite{Nalitov2012}. Averaging the product $\bm v_{\bm p}\bar{f}_{\bm p}(x)$ over directions of $\bm p$ and then, integrating by the electron energy, taking into account the oscillations of the density of states, we obtain the  ratchet current components in the form given in Appendix~\ref{App_theory}, Eq.~\eqref{j_tot_fin}. In general, we obtain that the leading contribution to the magnetic ratchet current is proportional to $\partial^2\delta_c/\partial 	\varepsilon_\text{F}^2 \approx (2\pi/\varepsilon_\text{F})^2\delta_c$. This  yields 1/$B$-oscillations which are in phase with the SdH oscillations.  Note that there is also a contribution proportional to the first derivative of $R_{xx}$ with respect to $B$ ($j \propto  dR_{xx}/dB$) phase-shifted by $\pi/2$ from $R_{xx}(B)$. However, this contribution is small with respect to the quantum parameter  $\hbar \omega_c/(2 \pi \varepsilon_{\text F}) \ll 1$, and, consequently, we omit it in the following calculations.

The obtained expressions, valid for arbitrary relation between the parameters $\omega_c$, $\omega$, and $\tau$, are cumbersome, therefore we give them here in two limits of low and high magnetic fields where $\omega_c$ is much higher and much smaller than $\omega$, respectively.

At $1/\tau \ll \omega_c \ll \omega$ (i.e. $B\ll B_{CR}$) we have: 
\begin{align}
\label{j_tot_low_B}
j_x= &  -{j_0 \over  (\omega_c\tau)^2} \left( {2\pi \varepsilon_\text{F}\over \hbar\omega_c}  \right)^2 \delta_c \\
&\times \left[1 -  {5\over 8(\omega_c\tau)^2} P_{\rm L} + {1\over 4\omega_c\tau} \tilde{P}_{\rm L} + {3\omega_c\over2\omega}  P_\text{circ}\right],\nonumber \\
j_y= &  {j_0 \over  2(\omega_c\tau)^2}\left( {2\pi \varepsilon_\text{F}\over \hbar\omega_c}  \right)^2 \delta_c \\
&\times \left[{4\over \omega_c\tau} +{1\over2\omega_c\tau} P_{\rm L} + {5\over 4(\omega_c\tau)^2} \tilde{P}_{\rm L} +  {5 \over \omega \tau} P_\text{circ}\right].\nonumber 
\end{align}

In the opposite limit $1/\tau \ll \omega \ll \omega_c$ ($B\gg B_{CR}$) we get:
\begin{align}
\label{j_tot}
j_x= &  -{j_0 \omega^2\over 4\omega_c^4\tau^2} \left( {2\pi \varepsilon_\text{F}\over \hbar\omega_c}  \right)^2 \delta_c \\
&\times \left[- 1 -  {7\over 2(\omega_c\tau)^2} P_{\rm L} + {3\over\omega_c\tau} \tilde{P}_{\rm L} +  {3\omega\over2\omega_c}  P_\text{circ}\right],\nonumber \\
\label{j_tot_y}
j_y= &  {j_0 \omega^2\over 2\omega_c^4\tau^2} \left( {2\pi \varepsilon_\text{F}\over \hbar\omega_c}  \right)^2 \delta_c \\
&\times \left[-{1\over 4\omega_c\tau} + {3\over2\omega_c\tau} P_{\rm L} + {7\over 4(\omega_c\tau)^2} \tilde{P}_{\rm L} + {1 \over \omega \tau} P_\text{circ}\right].\nonumber 
\end{align}
Here $j_0$ is the zero-field value of the ratchet current, Eq.~\eqref{j0}.

Above we developed the drift-diffusion theory assuming that the impurity scattering dominates over the electron-electron scattering. The discussion of the hydrodynamic regime,  where electron-electron collisions are very fast, will be presented elsewhere (previous studies on the hydrodynamic ratchet effect~\cite{Rozhansky2015,Popov2016,Fateev2017,Fateev2019} did not discuss magneto-oscillations). Our preliminary estimates show that similar results can be obtained. In particular, we find that in the high-field limit $\omega_c \gg \omega$ at $\bm E \parallel x$ the ratio $j_x/j_y \sim \omega_c\tau$ in accordance with Eqs.~\eqref{j_tot},~\eqref{j_tot_y}.

\section{Discussion}
\label{discussion}

Now we discuss the experimental results in the view of the developed theory. In the experiments, we probed the photocurrent flowing in the $x$ direction normal to the DGG stripes, because of the device geometry. The photocurrent obtained in Sec.~\ref{theory} is proportional to $\delta_c$, i.e., exhibits $1/B$-periodic oscillations  following SdH-oscillations. Analyzing the extrema positions of the photocurrent, see  Fig.~\ref{magnetic}, we obtained that the oscillations indeed correspond to the SdH oscillations of $R_{xx}(B)$, i.e. ratchet photocurrent  $J$ is proportional to $\delta_c$ as expected from Eqs.~\eqref{j_tot_low_B},~\eqref{j_tot}. We note that at high magnetic fields the experimentally observed ratchet current oscillations have a magnetic field dependent phase shift, and, hence, the photocurrent is phase-shifted in respect to the oscillations of $R_{xx}(B)$. Exploring this exciting feature requires additional experimental and theoretical studies, and is out of scope of this paper.

From Eqs.~\eqref{j_tot_low_B} and~\eqref{j_tot} together with Eq.~\eqref{delta_c} we expect sign-alternating  oscillations of the magnetic ratchet current as a function of Fermi energy and, consequently, gate voltages. The oscillations periodic in gate voltage are indeed observed at a fixed magnetic field for back- as well as top-gates, see Figs.~\ref{backgate} and~\ref{topgate}, respectively. Comparing Figs.~\ref{backgate} and \ref{topgate} we see a substantial difference in the period of oscillations. This is just caused by the different thicknesses (capacities) of the corresponding insulator layers. 

As a fingerprint of the  ratchet effect, the magneto-photocurrent $j_x \propto j_0$ is proportional to the lateral asymmetry parameter $\Xi$, see Eqs.~\eqref{zerocurrent} and~\eqref{Xi1}. The latter can easily be varied by the variation of the top gate polarities and relative amplitudes. This is indeed observed in experiment, see Fig.~\ref{topgate}, which shows that for large top gate voltages the sign of oscillations is opposite for opposite signs of the parameter $\Xi$, see black and red vertical arrows in Fig.~\ref{topgate}. Note that for both top gate voltages equal to zero the magnetic ratchet currents are caused by the built-in asymmetry, see Fig.~\ref{topgate}.

The developed theory also demonstrates that, as observed in the experiment, the magneto-oscillations of the ratchet current are highly sensitive to the  polarization state of the incident radiation. For linearly polarized radiation the magnetic ratchet current consists of the polarization-independent current as well as of two contributions varying upon rotation of the radiation polarization plane as $P_{\rm L}=\cos{2\alpha}$ and $\tilde{P}_{\rm L}= \sin{2\alpha}$,  see Eqs.~\eqref{j_tot_low_B},~\eqref{j_tot}. These contributions are clearly detected in experiment, see Fig.~\ref{alpha}, and  exhibit sign-alternating magneto-oscillations. The experimental results for the corresponding factors $J_D$ and $J_A$, yielding dominating contributions for wide range of magnetic fields,  are presented in Figs.~\ref{alpha}(a) and~\ref{magnetic}. Note that the photocurrent shown in the latter figure is obtained for $\alpha=0$ and presents the sum of $J_A$ and $J_D$.

The theory also describes well the dependences  of the oscillation amplitudes on the magnetic field. Figure~\ref{fig:theory} shows calculated magnetic field dependences of the coefficients $j_D$ and $j_A$ describing polarization-independent magnetic ratchet photocurrent density $j_D$, and the one driven by linearly polarized radiation, $j_A P_{\rm L}$. Note that the coefficients $j_D$ and $j_A$ are introduced in the same way as used in the experimental fit function Eq.~\eqref{j_alpha} having current contributions $J_D$ and $J_A P_{\rm L}(\alpha)$. The overall behavior of the oscillations is the same: an increase of magnetic field first giantly magnifies the oscillations magnitude, which, however, decreases for further magnetic field increase. Analytically, this non-monotonous behavior, which is absent in SdH oscillations, is described by the ratio of the current oscillation amplitude $j^*$ introduced according to $j_D={j^*} \cos{(\pi\varepsilon_\text{F}/\hbar\omega_c)}$ to that of the zero field current $j_0$. Then, for the polarization independent contribution, we obtain from Eq.~\eqref{j_tot_low_B} 
\begin{equation}
\label{amplitude}
\frac{{j^*}}{j_0} \sim \left({2\pi\varepsilon_\text{F} \over \hbar\omega_c}\right)^2 {\exp{(-\pi/\omega_c\tau_q)}\over (\omega_c\tau)^2}.
\end{equation}
A giant increase of the ratchet current when applying a magnetic field is caused by the first factor since $\varepsilon_F/\hbar\omega_c \gg 1$ so that $j^*>j_0$ up to $\omega_c\tau \approx 9$. Whereas for $\omega_c\tau \lesssim 2.2$ the amplitude $j^*$ rises due to factor $\exp(-\pi/\omega_c \tau_q)$ and $j^*/j_0 \gg 1$, a further increase in magnetic field leads to a decrease in the ratio. This is caused by the competition of the exponential factor, saturating at $\omega_c \tau_q \gg 1$, with the factor $\omega_c^{-4}$. This results in a maximum of the current, see Fig.~\ref{fig:theory}, clearly detected in experiment, see Fig.~\ref{alpha}(a). Note that a possible contribution of the Seebeck ratchet effect in quantizing magnetic fields~\cite{Budkin2016a,Faltermeier2017} can increase/decrease the magnitude of the polarization-independent current.

Similar analysis of the magnetic field dependence can also be performed  for the linear-polarization driven photocurrent $j_A$. From Eq.~\eqref{j_tot_low_B} follows that, alike $j_D$, the photocurrent $j_A$ drastically increases with the magnetic field increase, reaches maximum and decreases at further magnetic field increase. The only difference between the amplitudes  $j_A$ and $j_D$ is that in a high fields $j_A$ decreases faster, as $\omega_c^{-6}$. Figure~\ref{fig:theory}  shows that, for the magnetic field relevant to experiments, both amplitudes, $j_A$ and $j_D$,  yield  comparable contributions. This agrees with experiment, see Fig.~\ref{alpha}(a).  Furthermore, from  Eq.~\eqref{j_tot_low_B}  and  Fig.~\ref{fig:theory} it  follows that, at low magnetic fields, the polarization-independent component amplitude $j_D$ and that for the current sensitive to the orientation of the radiation electric field vector $j_A$ have opposite signs. This is in a fully agreement with experiment, see Fig.~\ref{alpha}(a) for $B \leq 2.5$~T. Note that more detailed comparison of the theory and experiments is complicated by  the magnetic field dependent phase shift addressed above as well as by a possible contribution of plasmonic effects~\cite{Rozhansky2015,Popov2016,Fateev2017,Fateev2019}.

Besides the photocurrent sensitive to the degree of linear polarization,  experiments show a substantial input of the magnetic ratchet current $j_x\propto P_{\text{circ}}$, which changes its sign upon switching the radiation helicity. Figures~\ref{circular} and~\ref{circular2}(a) show magneto-oscillations of this current, %($J_C P_\text{circ}$) 
and its dependence on the phase angle $\varphi$. The helicity-driven contribution is also expected from the developed theory, see last terms in square brackets on the right sides of Eqs.~\eqref{j_tot_low_B} and~\eqref{j_tot}. For circularly polarized radiation, the photocurrents proportional to $P_{\rm L}$ and $\tilde{P}_{\rm L}$ vanish, and the total ratchet current for high magnetic fields is given by ${j_x \propto 1 - (3\omega/2\omega_c)P_\text{circ}}$. For magnetic fields $B>\SI{4}{\tesla}$ and $f=\TeraHertz{2.54}$ we obtain that the amplitudes of the helicity-dependent and polarization-independent currents are comparable, which is in agreement with experiment, see Fig.~\ref{circular2}(a). Note that, in agreement with Eq.~\eqref{j_tot}, these currents have opposite signs. Similarly to the magnetic ratchet current driven by linearly polarized radiation, oscillations are expected as a function of top gate voltage and, indeed detected in the experiment, see Fig.~\ref{circular2}(b).

\section{Summary} 
\label{summary}

Our experiments together with the developed theory show that terahertz radiation applied to graphene with asymmetric, lateral superlattice and subjected to a strong magnetic field promotes the magnetic quantum ratchet effect. The characteristic feature of the magnetic field induced ratchet current is magneto-oscillations with a magnitude much larger than the ratchet current at zero magnetic field. This, caused by Shubnikov-de~Haas effect, enhancement of the ratchet effect is insofar generic as it is not only observed in graphene superlattices, but also in quantum well structures with parabolic spectrum~ \cite{Faltermeier2017,Faltermeier2018}. The amplitude and direction of the ratchet current are controlled by the lateral asymmetry parameter $\Xi$, magnetic field strength/direction, and the radiation's polarization state. The latter reflects magnetic ratchet current contributions driven by linearly and circularly polarized radiation. The presented theory describes well almost all results. It cannot, however, explain  the  magnetic field dependent phase shift of the ratchet current oscillations observed at high magnetic fields. This striking result may be caused by contributions from plasmonic ratchets~\cite{Rozhansky2015}, neglected here. Its understanding is a future task. 

To conclude, we observed  giant ratchet magneto-photocurrent in graphene lateral superlattices caused by  the Shubnikov-de Haas  effect  and developed a theory, which  explains  well experimental observations.

\section{Acknowledgments}
\label{acknow} The support from the FLAG-ERA program (project DeMeGRaS, DFG No. GA 501/16-1) and the Volkswagen Stiftung Program (97738) is gratefully acknowledged. The work of L.E.G. and V.Yu.K. was supported by the Foundation for the Advancement of Theoretical Physics and Mathematics ``BASIS''. L.E.G. also thanks Russian Foundation for Basic Research (project 19-02-00095). V.Yu.K . also thanks   Russian Foundation for Basic Research (Grant No. 20-02-00490). S.D.G. and V.Yu.K. thank Foundation for Polish Science (IRA Program, grant MAB/2018/9, CENTERA) for support.

\appendix

\section{Derivation of the ratchet current}
\label{App_theory}

The ratchet current Eq.~\eqref{j_tot} is obtained by sequential iterations of the kinetic equation~\eqref{kin_eq} at two small perturbations, namely the light amplitude $\bm E$ and the periodic ratchet potential $V(x)$. The first iteration step is always account for $\bm E$, but the next steps could be different. One contribution is obtained if the potential $V$ is taken into account at the second stage, and the radiation amplitude $\bm E$ at the last stage. We denote the corresponding correction to the distribution function $f^{(EVE)}$. In contrast to systems with parabolic energy dispersion, the total ratchet current is not restricted to this contribution. An additional contribution to the ratchet current, $\delta \bm j$, is obtained if the amplitude $\bm E$ is taken into account twice assuming $V=0$, and then, at the last stage, the periodic potential is taken into account. The corresponding part of the distribution function is denoted as $f^{(EEV)}$. In the next two subsections we derive both contributions to the ratchet current.

\subsection{$\bf EVE$ contribution}
\label{App_EVE}

%The ratchet current is given by
%\begin{equation}
%\label{j_def}
%j_\alpha = e\sum_{\bm p} v_\alpha f^{(EVE)},
%\end{equation}
%where $\bm v = \partial \varepsilon /\partial \bm p$ is the carrier velocity. 
The distribution function $f^{(EVE)}$ is a solution of the kinetic equation bilinear in $\bm E$ and linear in $V(x)$ obtained by a simultaneous account for $\bm E$, $V(x)$, and then $\bm E$. 
%(the other correction $\delta \bm j^{(EEV)}$ obtained by twice account for $\bm E$ and then for $V(x)$ is calculated in the next subsection).

The kinetic equation for $f^{(EVE)}$ has the form
\begin{equation}
\omega_c {\partial f^{(EVE)} \over \partial  \varphi} + e \bm E^* \cdot {\partial f^{(EV)} \over \partial \bm p} = - \gamma f^{(EVE)},
\end{equation}
where $f^{(EV)}$ is the correction bilinear in $\bm E$ and $V(x)$,
and 
\begin{equation}
\gamma(\varepsilon)={1\over\tau} \left[ 1+\delta_c(\varepsilon) \right].
\end{equation}

Solution is given by
\begin{equation}
\label{f_EVE}
f^{(EVE)} = - \sum_\pm \tau_c^\pm e \bm E^* \cdot \left( {\partial f^{(EV)} \over \partial \bm p} \right)_\pm + c.c.,
\end{equation}
where $(\ldots)_\pm$ denotes the $\pm 1$st Fourier-harmonics, and 
\begin{equation}
\tau_c^\pm = {1 \over \gamma(\varepsilon) \pm i\omega_c}.
\end{equation}
Here $\delta_c(\varepsilon)$ means $\delta_c$ given by Eq.~\eqref{delta_c} where $\varepsilon_\text{F}$ is substituted by $\varepsilon$.

Substituting the solution~\eqref{f_EVE} into Eq.~\eqref{j}, we obtain the current density in the form
\begin{equation}
j_\alpha = -e^2\sum_{\bm p} v_\alpha \sum_\pm \tau_c^\pm  \bm E^* \cdot \left( {\partial f^{(EV)} \over \partial \bm p} \right)_\pm + c.c.
\end{equation}
This equation shows that only the even in $\bm p$ part of $f^{(EV)}$ contributes to the photocurrent. It contains two terms, the $\varphi$-independent one and the 2nd harmonics of $\varphi$.
For $j_+=j_x+ij_y$ we get
\begin{equation}
j_+ = -e^2\sum_{\bm p} v_+  \tau_c^-  \bm E^* \cdot \left( {\partial f^{(EV)} \over \partial \bm p} \right)_- \\+ (\bm E \leftrightarrow \bm E^*, \:  \omega \rightarrow -\omega).
\end{equation}
Integrating by parts we obtain
\begin{equation}
j_+ = e^2\sum_{\bm p} {\partial (v_+  \tau_c^-)\over \partial \bm p}  \cdot  \bm E^* f^{(EV)} + (\bm E \leftrightarrow \bm E^*, \: \omega \to -\omega).
\end{equation}
Calculating the gradient in the momentum space
\begin{equation}
{\partial (v_+  \tau_c^-)\over \partial \bm p}  \cdot  \bm E^* 
= {v_0^2 \varepsilon \over 2} \left({\tau_c^-\over\varepsilon} \right)'  \left[(E^*)_+  + (E^*)_- \text{e}^{2i\varphi_{\bm p}} \right], 
\end{equation}
where $(E^*)_\pm = E^*_x\pm iE^*_y$,
and the prime denotes differentiating over $\varepsilon$,
we obtain
\begin{widetext}
	\begin{equation}
	j_+ = {e^2v_0^2\over 2} \sum_{\bm p} \varepsilon \left({\tau_c^-\over\varepsilon} \right)' \left[(E^*)_+  f^{(EV)}_0 + (E^*)_- f^{(EV)}_{-2} \right]  + (\bm E \leftrightarrow \bm E^*, \: \omega \to -\omega).
	\end{equation}
	Here $f^{(EV)}_{0,-2}$ mean the angular-independent part of $f^{(EV)}$ and the part $\propto \text{e}^{-2i\varphi_{\bm p}}$.

	Since the angular integration is already performed,  we can pass from summation over $\bm p$ to integration over energy:
	\begin{equation}
	\label{j+}
	j_+ = {e^2v_0^2\over 2} \int d\varepsilon g(\varepsilon) \varepsilon \left({\tau_c^-\over\varepsilon} \right)' \left[(E^*)_+  f^{(EV)}_0 + (E^*)_- f^{(EV)}_{-2} \right]  + (\bm E \leftrightarrow \bm E^*, \: \omega \to -\omega),
	\end{equation}
	where $g(\varepsilon)$ is the density of states.
\end{widetext}
The corrections $f^{(EV)}_{0,-2}$ are given by
\begin{equation}
f^{(EV)}_0 = {iev_0^2\over 4 \omega} \sum_\pm \left[-f_0' E_x{dV\over dx} {\left(\varepsilon \tau_{1\omega}^\pm  \right)' \over \varepsilon} + V{d E_x\over dx}  \tau_{1\omega}^\pm f_0''\right],
\nonumber
\end{equation}
\begin{equation}
f^{(EV)}_{-2} = {ev_0^2\tau_{2\omega}^-\over 4}  \left[-f_0' E_+{dV\over dx} \varepsilon \left({ \tau_{1\omega}^-   \over \varepsilon}\right)' + V{d E_+\over dx}  \tau_{1\omega}^- f_0''\right],
\nonumber
\end{equation}
where 
\begin{equation}
\label{tau_n_omega}
\tau_{n\omega}^\pm = {1\over \gamma(\varepsilon)-i\omega \pm in\omega_c},
\quad n=1,2.
%\tau_{1\omega}^\pm = {1\over \Gamma +\gamma(\varepsilon)-i\omega \pm i\omega_c},
%\qquad
%\tau_{2\omega}^\pm = {1\over \Gamma +\gamma(\varepsilon)-i\omega \pm 2i\omega_c}.
\end{equation}
These expressions at $\omega_c=0$ pass into the corresponding expressions from Ref.~\cite{Nalitov2012}.

Substituting $f^{(EV)}_{0,2}$ into Eq.~\eqref{j+}, we obtain 
\begin{equation}
\label{j+tot}
j_+ = j_+^{(0)} +j_+^{(-2)}+ (\bm E \leftrightarrow \bm E^*, \: \omega \to -\omega),
\end{equation}
where
\begin{widetext}
	%a contribution $j_+^{(0)}$ to the current:
	\begin{equation}
	j_+^{(0)} = {ie^3v_0^4\over 8\omega} \sum_\pm \int d\varepsilon g(\varepsilon) \varepsilon \left({\tau_c^-\over\varepsilon} \right)'  (E^*)_+  \left[-f_0' E_x{dV\over dx} {\left(\varepsilon \tau_{1\omega}^\pm  \right)' \over \varepsilon} + V{d E_x\over dx}  \tau_{1\omega}^\pm f_0''\right],
	\end{equation}
	\begin{equation}
	j_+^{(-2)} ={e^3v_0^4\over 8} \int d\varepsilon g(\varepsilon) \varepsilon\tau_{2\omega}^- \left({\tau_c^-\over\varepsilon} \right)'   (E^*)_-   \left[-f_0' E_+{dV\over dx} \varepsilon \left({ \tau_{1\omega}^-   \over \varepsilon}\right)' + V{d E_+\over dx}  \tau_{1\omega}^- f_0''\right].
	\end{equation}
	Averaging over the $x$ coordinate with $E_0$ being the near-field amplitude yields
	\begin{equation}
	\overline{E_0V{d E_0\over dx}} ={1\over 2}\overline{V{dE_0^2 \over dx}} =  -{1\over 2}\overline{E_0^2 {dV\over dx}} \equiv -{1\over 2}\Xi,
	\end{equation}
	and integrating over $\varepsilon$  we get
	\begin{equation}
	j_+^{(0)} = \Xi (|e_x|^2+ie_xe_y^*) {ie^3v_0^4\over 8\omega} 
	\sum_\pm \left\{ g   \left({\tau_c^-\over\varepsilon_\text{F}} \right)' \left(\varepsilon_\text{F} \tau_{1\omega}^\pm  \right)'
	-{1\over 2} \left[g  \varepsilon_\text{F} \left({\tau_c^-\over\varepsilon_\text{F}} \right)' \tau_{1\omega}^\pm \right]' \right\} ,
	\end{equation}
	\begin{equation}
	j_+^{(-2)} = \Xi (1-P_\text{circ}) {e^3v_0^4\over 8}  
	\left\{ g \varepsilon_\text{F}^2 \tau_{2\omega}^- \left({\tau_c^-\over\varepsilon_\text{F}} \right)'\left({ \tau_{1\omega}^-   \over \varepsilon_\text{F}}\right)'
	-{1\over 2} \left[g  \varepsilon_\text{F} \tau_{2\omega}^-\left({\tau_c^-\over\varepsilon_\text{F}} \right)' \tau_{1\omega}^- \right]' \right\}.
	\end{equation}
	Here the prime denotes differentiation over $\varepsilon_\text{F}$.
\end{widetext}

Let us analyze the terms in the curly brackets. The maximal result comes from the second derivative $(\tau_c^-)''$, therefore, only the second terms in curly brackets are important. The terms with the first derivative 
%are proportional to $\delta_s$, see Eq.~\eqref{d_s} of the main text, but their 
have much smaller amplitude due to the factor ${\hbar\omega_c/(2\pi\varepsilon_\text{F}) \ll 1}$. The terms $\sim \delta_c^2$ are also omitted because they have an additional small factor $\exp(-\pi/\omega_c\tau_q)\ll 1$ and result in oscillations with double period not present in the experiment. As a result, we obtain
\begin{equation}
j_+^{(0)} = -\Xi (i|e_x|^2-e_xe_y^*) {e^3v_0^4 g\over 8} 
%{1\over 2\omega} \sum_\pm \tau_{1\omega}^\pm
T_\omega (\tau_c^-)'' ,
\end{equation}
\begin{equation}
j_+^{(-2)} = - \Xi (1-P_\text{circ}) {e^3v_0^4g\over 16}  
%\tau_{2\omega}^- \tau_{1\omega}^-  
Q_\omega (\tau_c^-)'',
\end{equation}
where ${g=2\varepsilon_\text{F}/(\pi\hbar^2v_0^2)}$ is the zero-field density of states (spin and valley degeneracies are taken into account), and
\begin{equation}
T_\omega = {\tau_{1\omega}^+ + \tau_{1\omega}^-\over 2\omega},
%{1\over 2\omega} \sum_\pm \tau_{1\omega}^\pm
\qquad
Q_\omega = \tau_{2\omega}^- \tau_{1\omega}^-.
\end{equation}
%\begin{equation}
%j_+^{(0)} = -{1\over 2} \Xi (|e_x|^2+ie_xe_y^*) {ie^3v_0^4\over 8\omega} 
%\sum_\pm \tau_{1\omega}^\pm
%%\left[ -g'{\tau_c^- \over \varepsilon_\text{F}} + 
%g (\tau_c^-)'' 
%%\right] 
%,
%\end{equation}
%\begin{equation}
%j_+^{(-2)} = -{1\over 2} \Xi (1-P_\text{circ}) {e^3v_0^4\over 8}  
%%\left[-g'  {\tau_{2\omega}^-\tau_c^- \over\varepsilon_\text{F}} \tau_{1\omega}^- + 
%g \tau_{2\omega}^- (\tau_c^-)'' \tau_{1\omega}^- 
%%\right]
%.
%\end{equation}

Finally, from Eq.~\eqref{j+tot} we obtain
the total current
\begin{align}
\label{j_fin}
j_+  & = -\Xi{e^3 v_0^2 \varepsilon_\text{F} \over 4\pi \hbar^2 } (\tau_c^-)''\\
%\left( {4\pi \varepsilon_\text{F}\over \hbar\omega_c} \right)^2 \delta_c  
&\times \left[Q_+ + iT_+(1+P_{\rm L}+i\tilde{P}_{\rm L})
+ (iT_- - Q_-)P_\text{circ}  \right] 
%%%%%%%   
.\nonumber
\end{align}
Here $T_\pm=(T_\omega \pm T_{-\omega})/2$, $Q_\pm=(Q_\omega \pm Q_{-\omega})/2$.
%$P_{\rm L}=2|e_x|^2-1=|e_x|^2-|e_y|^2$, $\tilde{P}_L=e_xe_y^*+e_x^*e_y$.

%Comparing with the zero-field value of the linear ratchet current Eq.~\eqref{zerocurrent}
%we get
%\begin{multline}
%\label{j_fin}
%\begin{pmatrix}
%j_x \\ j_y 
%\end{pmatrix}=
%j_0  {4\pi \varepsilon_\text{F}\over \hbar\omega_c} \left( 
%\delta_s
%- {\gamma\over\Gamma}{4\pi \varepsilon_\text{F}\over \hbar\omega_c}\delta_c
%\right) \\ 
%\times
%\begin{pmatrix}
%1 & {\omega_c\over\omega} & -{\Gamma\over\omega} & 0 \\
%0 & -{\Gamma\over\omega} & -{4\omega_c\over\omega} &1
%\end{pmatrix}
%\begin{pmatrix}
%P_{\rm L} \\ P_{\rm circ} \\ 1 \\ \tilde{P}_L 
%\end{pmatrix}.
%\end{multline}
%The obtained result is even in $B$, therefore it has the same symmetry as in zero field: the polarization-independent and $\propto \cos{2\alpha}$ signals are allowed in the $x$ direction while the circular and $\propto \sin{2\alpha}$ currents flow in the $y$ direction.
%The current $j_x\propto |e_x|^2=(1+\cos{2\alpha})/2$ can be obtained from $j_+^{(0)}$ in the next order in the small parameter $\omega/\Gamma$ (the odd in $\omega_c/\Gamma$ terms cancel each other).

\subsection{$\bf EEV$ contribution}
\label{App_EEV}

Here we calculate the  correction $\delta \bm j$ obtained by twice account for $\bm E$ and then for $V(x)$.
The corresponding ratchet current is given by
\begin{equation}
\label{j_def}
\delta j_\alpha = e\sum_{\bm p} v_\alpha f^{(EEV)}.
\end{equation}
%where $\bm v = \partial \varepsilon /\partial \bm p$ is the carrier velocity. The distribution function $f^{(EVE)}$ is a solution of the kinetic equation bilinear in $\bm E$ and linear in $V(x)$.

The kinetic equation for $f^{(EEV)}$ has the form
\begin{equation}
\omega_c {\partial f^{(EEV)} \over \partial  \varphi} - {dV\over dx} {\partial f^{(EE)} \over \partial p_x} = - \gamma f^{(EEV)},
\end{equation}
where $f^{(EE)}$ is the correction bilinear in $\bm E$.
Solution is given by
\begin{equation}
\label{f_EEV}
f^{(EEV)} = \sum_\pm \tau_c^\pm {dV\over dx} \left( {\partial f^{(EE)} \over \partial p_x} \right)_\pm.
\end{equation}
%where $(\ldots)_\pm$ denotes the $\pm 1$st Fourier-harmonics, and 
%\begin{equation}
%\tau_c^\pm = {1 \over \Gamma +\gamma(\varepsilon) \pm i\omega_c}.
%\end{equation}

Substituting the solution~\eqref{f_EEV} into Eq.~\eqref{j_def}, we obtain the current density in the form
\begin{equation}
\delta j_\alpha = e{dV\over dx} \sum_{\bm p} v_\alpha \sum_\pm \tau_c^\pm \left( {\partial f^{(EE)} \over \partial p_x} \right)_\pm.
\end{equation}
%This equation shows that only the even in $\bm p$ part of $f^{(EV)}$ contributes to the photocurrent. It contains two parts, the $\varphi$-independent and the 2nd harmonics of $\varphi$.

For $\delta j_+=\delta j_x+i\delta j_y$ we get
integrating by parts
\begin{equation}
\delta j_+ = -e{dV\over dx} \sum_{\bm p} {\partial (v_+  \tau_c^-)\over \partial p_x}  f^{(EE)}.
\end{equation}

Calculating the derivative
\begin{equation}
{\partial (v_+  \tau_c^-)\over \partial p_x}   
= {v_0^2 \varepsilon \over 2} \left({\tau_c^-\over\varepsilon} \right)'  \left(1  +  \text{e}^{2i\varphi_{\bm p}} \right), 
\end{equation}
we obtain
\begin{equation}
\delta j_+ = -{e v_0^2\over 2}{dV\over dx} \sum_{\bm p} \varepsilon \left({\tau_c^-\over\varepsilon} \right)' \left(f^{(EE)}_0 + f^{(EE)}_{-2} \right) .
\end{equation}
Here $f^{(EE)}_{0,-2}$ mean the angular-independent part of $f^{(EE)}$ and the part $\propto \text{e}^{-2i\varphi_{\bm p}}$. The former is controlled by energy relaxation processes: $f^{(EE)}_0\propto \tau_\varepsilon$ with $\tau_\varepsilon$ being the energy relaxation time. It describes the Seebeck and Nernst-Ettingshausen ratchet effects~\cite{Budkin2016a}. In what follows we omit this contribution concentrating on polarization-dependent ratchet currents.

Since the angular integration is already performed,  we can pass from summation over $\bm p$ to integration over energy:
\begin{equation}
\label{d_j+}
\delta j_+ = -{ev_0^2\over 2}{dV\over dx} \int d\varepsilon g(\varepsilon) \varepsilon \left({\tau_c^-\over\varepsilon} \right)'  f^{(EE)}_{-2} ,
\end{equation}
where $g(\varepsilon)$ is the density of states.
The correction  $f^{(EE)}_{-2}$ is multiplied by $dV/dx$, therefore we find it in the quasi-homogeneous limit:
\begin{equation}
-2i\omega_c f^{(EE)}_{-2} + e\bm E^*\cdot \left( {\partial f^{(E)} \over \partial \bm p} \right)_{-2} = - \gamma f^{(EE)}_{-2}, 
\end{equation}
where the linear in $\bm E$ correction to the distribution function is found from
\begin{equation}
\omega_c {\partial f^{(E)} \over \partial  \varphi}+ e\bm E\cdot\bm v f_0' = - {f^{(E)}\over \tau_{1\omega}}.
\end{equation}
The solutions are:
\begin{equation}
f^{(E)} 
%= -f_0' e\tau_{\omega}^+ \bm E\cdot\bm v 
= -{ev_0\over 2}\sum_\pm f_0' \tau_{1\omega}^\pm {p_\pm\over p} E_\mp,
\end{equation}
\begin{equation}
f^{(EE)}_{-2} = - \tau_{c2} e\bm E^*\cdot \left( {\partial f^{(E)} \over \partial \bm p}\right)_{-2} + (\bm E \leftrightarrow \bm E^*, \: \omega \to -\omega),
\end{equation}
where $\tau_{1\omega}^\pm$ is given by Eq.~\eqref{tau_n_omega}, and
\begin{equation}
%\tau_{\omega}^\pm = {1\over \Gamma +\gamma(\varepsilon)-i\omega \pm i\omega_c},
%\qquad
\tau_{c2}  = {1\over \gamma(\varepsilon)-2i\omega_c}.
\end{equation}
%These expressions at $\omega_c=0$ pass  into the corresponding expressions from Ref.~\cite{Nalitov}.
Calculation is performed as follows:
\begin{equation}
\bm E^*\cdot \left( {\partial   \over \partial \bm p}f_0' \tau_{1\omega}^\pm {p_\pm E_\mp\over p}\right)_{-2}
%=(E^*)_+E_+  \left( {\tau_{\omega}^- f_0'\over p} \right)' {v_0p\over 2} 
= |E|^2(P_{\rm L}+i\tilde{P}_{\rm L}){v_0\over 2} \varepsilon \left( {\tau_{1\omega}^- f_0'\over \varepsilon} \right)' ,
\end{equation}
which yields
\begin{equation}
f^{(EE)}_{-2} = {e^2\tau_{c2}v_0^2\over 4}  |E|^2(P_{\rm L}+i\tilde{P}_{\rm L})\varepsilon \left[ {(\tau_{1\omega}^- + \tau_{1,-\omega}^-)f_0'\over \varepsilon} \right]'.
\end{equation}

Substituting $f^{(EE)}_{-2}$ into Eq.~\eqref{d_j+} and averaging over the $x$ coordinate, we obtain 
\begin{widetext}
	\begin{equation}
	\delta j_+  =-{e^3v_0^4\over 4}\Xi(P_{\rm L}+i\tilde{P}_{\rm L})\int d\varepsilon g(\varepsilon) \varepsilon^2 \left({\tau_c^-\over\varepsilon} \right)'   \tau_{c2}   \left[{(\tau_{1\omega}^- + \tau_{1,-\omega}^-)\over 2} {f_0'\over \varepsilon} \right]'.
	\end{equation}
\end{widetext}
Integrating over $\varepsilon$  we get
\begin{equation}
\delta j_+  =-{e^3v_0^4\over 4}\Xi(P_{\rm L}+i\tilde{P}_{\rm L}) \left[ g \varepsilon_\text{F}^2 \left({\tau_c^-\over\varepsilon_\text{F}} \right)' \tau_{c2} \right]'{\tau_{1\omega}^- + \tau_{1,-\omega}^-\over 2\varepsilon_\text{F}}.
\end{equation}

According to the same arguments as at calculation of the $EEV$-contribution (see the previous susbsection), the maximal result comes from 
%$g'$ and 
$(\tau_c^-)''$:
\begin{equation}
\label{d_j_fin}
\delta j_+  =-\Xi{e^3 v_0^2 \varepsilon_\text{F} \over 4\pi \hbar^2 } (\tau_c^-)''R(P_{\rm L}+i\tilde{P}_{\rm L}).
\end{equation}
%\begin{equation}
%\delta j_+  =-{e^3v_0^4\over 4}\Xi(P_{\rm L}+i\tilde{P}_L){(\tau_{1\omega}^- + \tau_{1,-\omega}^-)\tau_{c2}\over 2}    
%%\left[- g' \tau_c^- +  
%g (\tau_c^-)''. 
%% \right].
%\end{equation}
%
Here
\begin{equation}
R = \tau_{c2}(\tau_{1\omega}^- + \tau_{1,-\omega}^-).
\end{equation}

%Since 
%\begin{equation}
%\tau_{\omega}^- \approx {1\over \Gamma} +i {\omega+\omega_c\over\Gamma^2},
%\qquad
%\tau_{c2}\approx {1\over \Gamma} +i {2\omega_c\over\Gamma^2}
%\end{equation}
%we have 
%\begin{equation}
%{\tau_{\omega}^- + \tau_{-\omega}^-\over 2}\tau_{c2} 
%\approx {1\over \Gamma^2} \left({1\over \Gamma} + {3i\omega_c\over\Gamma} \right).
%\end{equation}
%Then substituting $\tau_c^-$, $g'$ and $(\tau_c^-)''$ from Eq.~\eqref{proizv},
%we obtain
%\begin{multline}
%\delta j_+  = -{e^3v_0^2\over 2 \pi\hbar^2\Gamma^3}\Xi(P_{\rm L}+i\tilde{P}_L)\left(1+ {4i\omega_c\over\Gamma} \right) \\ \times {4\pi \over \hbar\omega_c}   \left(\delta_s - {\gamma\over\Gamma}{4\pi \varepsilon_\text{F}\over \hbar\omega_c}\delta_c  \right).
%\end{multline}

%Finally we get
%\begin{multline}
%\label{d_j_fin}
%\begin{pmatrix}
%\delta j_x \\ \delta j_y 
%\end{pmatrix}=
%-2j_0^\text{lin} {4\pi \varepsilon_\text{F}\over \hbar\omega_c} \left( 
%\delta_s
%- {\gamma\over\Gamma}{4\pi \varepsilon_\text{F}\over \hbar\omega_c}\delta_c
%\right) \\ 
%\times
%\begin{pmatrix}
%1 & -{4\omega_c\over\Gamma} \\
%{4\omega_c\over\Gamma} & 1
%\end{pmatrix}
%\begin{pmatrix}
%P_{\rm L} \\ \tilde{P}_L 
%\end{pmatrix},
%\end{multline}
%where the zero-field value of the linear ratchet current $j_0$ is given by Eq.~\eqref{j0}.

\subsection{Total ratchet current}
\label{App_total}

A sum of $j_+$ from Eq.~\eqref{j_fin} and $\delta j_+$ from Eq.~\eqref{d_j_fin} yields the total ratchet current in the form
\begin{align}
\label{j_tot_fin}
j_x+ij_y  =  & -\Xi{e^3 v_0^2 \varepsilon_\text{F} \over 4\pi \hbar^2 } (\tau_c^-)''
\biggl[Q_+ + iT_+ \\ &  + (iT_+ + R)(P_{\rm L}+i\tilde{P}_{\rm L})
+ (iT_- - Q_-)P_\text{circ}  \biggr] 
.\nonumber
\end{align}
Substituting 
\begin{equation}
(\tau_c^-)''=\delta_c \left( {2\pi \over \hbar\omega_c} \right)^2    {\tau\over(1-i\omega_c\tau)^2},
\end{equation}
and finding real and imaginary parts of Eq.~\eqref{j_tot_fin}, we obtain the components of the total ratchet current. 

For two limiting cases of low and high magnetic fields, passing to the limits $\omega\tau \gg \omega_c\tau \gg 1$ and $\omega_c\tau \gg \omega\tau \gg 1$ we get Eqs.~\eqref{j_tot_low_B} and~\eqref{j_tot} of the main text, respectively.

%\FloatBarrier
%\bibliography{ref}
\input{ratchet_myfile4april.bbl}
\end{document}